 \documentclass[12pt,preprint]{aastex}
\usepackage{amsfonts}

\newcommand{\be}{\begin{equation}}
\newcommand{\ee}{\end{equation}} 
\newcommand{\bea}{\begin{eqnarray}}
\newcommand{\eea}{\end{eqnarray}}

\newcommand{\uvr}{\hat{r}}
\newcommand{\uvz}{\hat{z}}
\newcommand{\uvp}{\hat{\phi}}
\newcommand{\omegat}{\tilde{\omega}}

\begin{document}

\title{Do Magnetic Fields Destroy Black Hole Accretion Disk 
g-Modes?}

\author{Manuel Ortega-Rodr\1guez,\altaffilmark{1,2,3}
Hugo Sol\1s-S\'anchez,\altaffilmark{2}
J.~Agust\1n Arguedas-Leiva,\altaffilmark{2}
}
\affil{Escuela de F\1sica \& 
Centro de Investigaciones Geof\'{\i}sicas, 
Universidad de Costa Rica,\\ 11501-2060 San Jos\'e, Costa Rica}
\altaffiltext{1}{Visiting Scholar, KIPAC, Stanford University, 
Stanford, CA 94305-4060}
\altaffiltext{2}{Instituto de F\'{\i}sica Te\'orica, 
1248-2050 San Jos\'e, Costa Rica}
\altaffiltext{3}{Author to whom correspondence
should be addressed, {\tt manuel.ortega@ucr.ac.cr}}
\and
\author{Robert V. Wagoner, Adam Levine}
\affil{Department of Physics and Kavli Institute for Particle Astrophysics and Cosmology, Stanford University, Stanford, CA 94305-4060, USA}

\begin{abstract}
Diskoseismology, the theoretical study of normal mode oscillations
in geometrically thin, optically thick accretion disks,
is a strong candidate to explain some QPOs in the power spectra
of many black hole X-ray binary systems.
The existence of g-modes, presumably the most robust and visible of
the modes, 
depends
on general relativistic gravitational trapping in the hottest part of the disk.
As the existence of the required cavity 
in the presence of magnetic fields has been put into
doubt by theoretical calculations,
we will explore in greater generality what the inclusion of magnetic
fields has to say on the existence of g-modes. 
We use an analytical perturbative approach on the equations of MHD to
assess the impact of such effects.
Our main conclusion is that
there appears to be no compelling reason to discard g-modes.
In particular,
the inclusion of a non-zero {\it radial} component of the magnetic
field enables a broader scenario
for cavity non-destruction, 
especially taking into account
recent simulations' saturation values
for the magnetic field.
\end{abstract}


\section{INTRODUCTION}

There currently exists a rich structure
in the power spectra observations of black hole X-ray binary systems,
which includes
high frequency (40--450 Hz) quasi-periodic oscillations (QPO).
Relativistic diskoseismology,
the formalism of normal-mode oscillations of 
geometrically thin, optically thick accretion disks,
is a strong candidate to explain at least some of these QPOs.
(For a review, see Wagoner 2008.)

Diskoseismology's perturbative approach 
assumes that the effects of magnetic
fields have been incorporated in the background equilibrium solution
and works with fluid perturbations in which magnetic fields
play no effective role.
The objective of this paper is to study analytically the 
effects of
including small magnetic fields on  
the oscillations described by 
relativistic diskoseismology. 

Building on previous work (Fu \& Lai 2009, hereafter FL), 
we use a local (WKB) analysis of the full MHD equations to examine
how the magnetic field affects the physics of radial wave propagation.
The main difference with FL is that
we include all three components of the
magnetic field, not just the
vertical 
and toroidal cases separately.
We do assume, however, that the toroidal magnetic field 
component $B_\phi$ is larger than the other components
(using cylindrical coordinates $r$, $\phi$, $z$).
This assumption is supported by simulations (see Table 1).

We show that diskoseismic g-modes are more resistant to
magnetic-field disruption than previously thought.

\section{RELATIVISTIC DISKOSEISMOLOGY AND GRAVITATIONAL TRAPPING}

Within diskoseismology, some 
observed high-frequency oscillations 
in the outgoing radiation of black hole 
X-ray binary systems such as GRO J1655-40
are due to normal modes of  
adiabatic hydrodynamic perturbations. These modes  
are the result of 
gravitational driving and pressure restoring forces in 
a geometrically thin, optically thick accretion disc
in the weak thermal, steep power-law state
(Remillard \& McClintock 2006).

The study of diskoseismology reveals the existence
of different types of oscillation modes.
Of these, the fundamental g-mode
(an axisymmetric inertial-gravitational mode that oscillates mainly in 
the vertical plane) is the strongest 
candidate for explaining 
one of the QPOs, 
being the most robust and observable: 
it lies in the hottest part of the disk, 
has the largest photosphere, 
and is located
away from the uncertain physics of the inner boundary 
(Perez et al.~1997).  
The p-modes, on the other hand, are less observable.
They are only weakly affected by the magnetic fields (FL).

This interpretation is not only
supported
observationally by peaks in the 
power spectral density, but
the g-mode has been observed in 
hydrodynamic simulations as well (Reynolds \& Miller 2009;
O'Neill, Reynolds, \& Miller 2009). 

As one can see in Fig.~1, the fundamental g-mode
is trapped just
under the maximum value of the radial epicyclic frequency $\kappa(r)$
in the absence of magnetic fields.
Therefore, the existence of g-modes would be compromised by physical
conditions
that have an effect on the corresponding trapping curve. 
(The explanation of the other curves in the figure 
can be found in Section 4.) 

\section{THE EFFECTS OF INCLUDING MAGNETIC FIELDS}

The inclusion of magnetic fields modifies the shape of the trapping curve
in Fig.~1 and therefore threatens the existence of g-modes,
given that it is not certain whether the inner boundary
(the innermost stable circular orbit) can effectively trap
this type of mode.

\subsection{MHD Equations}

The Newtonian ideal MHD equations for non-self-gravitating accretion disks are:
\be\label{cont}
\frac{\partial \rho}{\partial t} + 
\nabla \cdot (\rho \mathbf{v}) = 0  \, ,
\ee
\be\label{mom}
\frac{\partial \mathbf{v}}{\partial t}
+ (\mathbf{v} \cdot \nabla)\mathbf{v}=
-\frac{1}{\rho}\nabla\Pi - \nabla\Phi + 
\frac{1}{\rho}\mathbf{T}  \, ,
\ee
\be\label{mhd}
  \frac{\partial \mathbf{B}}{\partial t}
  = \nabla \times (\mathbf{v} \times \mathbf{B})  \, ,
\ee
with
\be
  \Pi \equiv P + \frac{B^2}{8\pi} \, ,  \qquad 
  \mathbf{T} \equiv \frac{1}{4\pi}
  (\mathbf{B} \cdot \nabla)\mathbf{B}  \, .
\ee
These are seven equations for seven unknowns: 
the density $\rho$, the velocity $\mathbf{v}$, and the
magnetic field $\mathbf{B}$.
A barotropic pressure $P = P(\rho)$ is assumed, 
$\Phi$ is the pseudo-Newtonian gravitational potential, and of course
\be\label{div}
\nabla \cdot \mathbf{B} = 0 \, .
\ee

We restrict our analysis to standard thin disks.
We include the most important effects of general relativity by using
the exact expressions for the orbital angular velocity $\Omega(r)$ and
the radial epicyclic frequency $\kappa(r)$:
\bea
\Omega(r) &=& (r^{3/2}+a)^{-1} \, , \\
\kappa(r) &=& \Omega(r)(1-6/r+8a/r^{3/2}-3a^2/r^2)^{1/2} \, .
\eea
There is then no need to specify $\Phi(r,z)$.

The standard approach includes the assumptions
that the unperturbed background flow
be axisymmetric, 
with $\mathbf{v} = r \Omega(r) \uvp$, 
and that $\mathbf{B} = B_\phi(r) \uvp + B_z \uvz$,
where $B_z$ is a constant.
These forms of $\mathbf{v}$ and $\mathbf{B}$ satisfy
the stationary $(\partial/\partial t = 0)$ equations
(\ref{cont}) and (\ref{mhd}).
The radial force balance is dominated by the centrifugal
and the gravitational terms,
while the vertical balance is dominated by the vertical pressure gradient 
and the gravitational term. In other words, the magnetic fields are
non-dominant.

\subsection{Inclusion of Radial Magnetic Fields}

In order to extend the usual approach, we
include small (compared to $B_\phi$) radial magnetic fields.
The traditional exclusion of radial magnetic fields 
in analytical treatments has no justification
other than aesthetic prejudice (since they destroy 
the solutions' stationarity). 
In fact, simulations consistently yield
radial fields which are larger than vertical fields (see Table 1). 

We now discuss what happens in the formalism once one 
introduces a radial $B_r \uvr$ term in the magnetic field of
an otherwise stationary system, in particular the one described in the previous subsection.
If this term has the form $B_r = C/r$, where $C$ is a constant, 
then (\ref{div}) is immediately satisfied, while (\ref{mhd})
yields:
\be\label{newbphi}
   B_\phi^{\rm new} 
= B_\phi^{\rm old} + \Delta B_\phi(r,t) \, ,
\ee
where
\be
  \Delta B_\phi(r,t) \equiv r B_r \, \frac{d\Omega}{dr} \, t \, . 
\ee
We note immediately that $\Delta B(r,t) \sim B_r \ll B_\phi$ for
timescales of a few fluid oscillations 
(such that $\Omega t \sim 1$).

The new $B_\phi$ field satisfies equation (\ref{div}), and
(being parallel to $\mathbf{v}$)
yields 0 on the right-hand side of equation (\ref{mhd}).

In the context of the present paper on perturbations, we take $v_r$ as
negligible.  
However, it is important to verify that such an assumption does not lead to inconsistencies at the unperturbed (equilibrium) level. 
The important point is that once one accepts a non-vanishing $B_r$, 
then one must allow for a non-zero $v_r$ in order
to preserve the equality on equation 
(\ref{mom}); in particular, this equation acquires a $\phi$-component.

Fortunately, the implied value for $v_r \sim \varepsilon^2 \, b_r^2 \, r \Omega$
is in fact no larger than the one expected from
standard (Novikov \& Thorne 1973) 
viscosity considerations alone,
which is $v_r \sim \varepsilon^2 \alpha_* r \Omega$.
Here, $\alpha_*$ is the ``viscosity parameter'' (shear stress/pressure),
\be
 \varepsilon \equiv \frac{h}{R} = 
     \frac{\rm disk \ thickness}
    {{\rm typical \ value \ of \ } r} \, , 
\ee
while
$b_r$ is defined in equation (\ref{defbi}).

\subsection{Perturbations} 

We consider perturbations of 
$\delta \rho$, $\delta \mathbf{v}$, and $\delta \mathbf{B} 
\propto e^{im\phi-i\omega t}$.
Recall that our unperturbed magnetic field has the form
\be
 \mathbf{B} = B_\phi(r) \uvp + (C/r) \uvr + B_z \uvz \, .
\ee
We work with the assumptions
\be
B_r^2, B_z^2 \ll B_\phi^2 \ll 4\pi P \sim 4\pi\rho c_s^2 \,
\ee
(where $c_s$ refers to the speed of sound)
and define the following
small parameter:
\be\label{defbi}
  b_i \equiv \frac{B_i}{\sqrt{4\pi\rho c_s^2}} = \frac{v_{Ai}}{c_s} \, \qquad
  (i = r, z, \phi)  \, ,
\ee
where $\mathbf{v}_{\! A}$ is the Alfv\'en velocity.
The linearized equations for the perturbations then become:
\be\label{pcont}
-i \omegat \delta\rho 
+ \frac{1}{r} \frac{\partial}{\partial r} (\rho r \delta v_r)
+ \frac{im\rho}{r} \delta v_\phi
+ \frac{\partial}{\partial z} (\rho \delta v_z) = 0 \, ,
\ee
\be\label{pmomr}
-i \omegat \delta v_r - 2\Omega \delta v_\phi
= G_r \frac{\delta\rho}{\rho}
-\frac{1}{\rho} \frac{\partial}{\partial r} \delta \Pi \\
+\frac{1}{4\pi\rho} \left[
\frac{im}{r} B_\phi + B_z \frac{\partial}{\partial z}
{+ B_r \frac{\partial}{\partial r}
+ \frac{\partial B_r}{\partial r}
}
\right] \delta B_r
-\frac{B_\phi}{2\pi\rho r} \delta B_\phi  \, ,
\ee
\be\label{pmomp}
-i \omegat \delta v_\phi + \frac{\kappa^2}{2\Omega} \delta v_r
= 
{G_\phi \frac{\delta\rho}{\rho}}
-\frac{i m}{\rho r} \delta \Pi
+\frac{1}{4\pi\rho} \left[
\frac{im}{r} B_\phi + B_z \frac{\partial}{\partial z}
{+ B_r \left( \frac{\partial}{\partial r}
+ \frac{1}{r}
\right)}
\right] \delta B_\phi
+\frac{1}{4\pi\rho} \left( \frac{B_\phi}{r} 
 + \frac{\partial B_\phi}{\partial r}
\right) \delta B_r  \, ,
\ee
\be\label{pmomz}
-i \omegat \delta v_z = G_z \frac{\delta\rho}{\rho}
-\frac{1}{\rho} \frac{\partial}{\partial z} \delta \Pi
+\frac{1}{4\pi\rho} \left[
\frac{im}{r} B_\phi + B_z \frac{\partial}{\partial z}
{+ B_r \frac{\partial}{\partial r}}
\right] \delta B_z \, ,
\ee
\be\label{pmhdr}
  -i \omegat \delta B_r = \left(
\frac{i m B_\phi}{r} + B_z \frac{\partial}{\partial z}
\right) \delta v_r
{- \frac{i m B_r}{r} \delta v_\phi
- B_r \frac{\partial}{\partial z} \delta v_z} \, ,
\ee
\be\label{pmhdp}
  -i \omegat \delta B_\phi =
- \frac{\partial}{\partial r} (B_\phi \delta v_r) 
+ B_z \frac{\partial}{\partial z} \delta v_\phi 
- B_\phi \frac{\partial}{\partial z} \delta v_z
{
+ \frac{\partial}{\partial r} (B_r \delta v_\phi) 
} 
+ r \frac{d\Omega}{dr} \delta B_r \, ,
\ee
\be\label{pmhdz}
  -i \omegat \delta B_z = 
-\frac{B_z}{r} \frac{\partial}{\partial r} 
(r \delta v_r)
-\frac{im B_z}{r} \delta v_\phi
+\left(
\frac{im B_\phi}{r} 
{+ B_r \frac{\partial}{\partial r}}
\right) \delta v_z \, .
\ee
We have used the definitions
$\omegat \equiv \omega - m \Omega$ and
\be\label{g}
  \mathbf{G} \equiv \frac{1}{\rho}\nabla\Pi - \frac{1}{\rho}
\mathbf{T} \, .
\ee

\subsection{WKB Analysis of Axisymmetric Oscillations}

We now restrict ourselves to WKB conditions,
in which by definition 
all perturbations are $\propto e^{ik_r r + i k_z z}$.
Furthermore, we study for simplicity axisymmetric (i.e., $m=0$)
oscillations,
which are also more observationally relevant.
We use $\delta \Pi$ instead of $\delta \rho$.
Assuming $k_r, k_z \gg 1/r$ and $B_\phi \propto r^q$,
and using the definition
$p \equiv d \ln \Omega/d \ln r$, the following
equations obtain:
\be\label{wkbconta}
 -\frac{i\omega}{\rho c_s^2} \delta\Pi
+ i k_r \delta v_r  
+ i k_z \delta v_z 
+ \frac{i\omega}{4 \pi\rho c_s^2} 
(B_\phi \delta B_\phi 
{+ B_z \delta B_z} 
{+ B_r \delta B_r}) = 0 \, ,
\ee
\be\label{wkbmomra}
-\frac{ik_r}{\rho}\delta \Pi
+i \omega \delta v_r + 2\Omega \delta v_\phi
+\frac{1}{4\pi\rho} \left(
{i k_z B_z}
{+ i k_r B_r
}
\right) \delta B_r
-\frac{B_\phi}{2\pi\rho r} \delta B_\phi = 0 \, ,
\ee 
\be\label{wkbmompa}
i \omega \delta v_\phi - \frac{\kappa^2}{2\Omega} \delta v_r
+\frac{1}{4\pi\rho} 
\left(
{i k_z B_z}
{+ i k_r B_r
}
\right)
 \delta B_\phi
+\frac{(1+q) B_\phi}{4\pi\rho r} \delta B_r
= 0 \, ,
\ee 
\be\label{wkbmomza} 
-\frac{ik_z}{\rho}\delta\Pi +i \omega \delta v_z
+\frac{1}{4\pi\rho} 
\left(
{i k_z B_z}
{+ i k_r B_r
}
\right)
 \delta B_z 
= 0 \, ,
\ee  
\be\label{wkbmhdra}
i k_z B_z
\delta v_r + i \omega \delta B_r
{- i k_z B_r \delta v_z}
= 0 \, ,
\ee  
\be\label{wkbmhdpa}
i k_r B_\phi \delta v_r -
\left(
{i k_z B_z}
{+ i k_r B_r
}
\right)
 \delta v_\phi
+ i k_z B_\phi \delta v_z 
- i \omega \delta B_\phi - p \Omega \delta B_r = 0 \, ,
\ee
\be\label{wkbmhdza}
{i k_r B_z \delta v_r} 
{- i k_r B_r \delta v_z} - i \omega \delta B_z
= 0 \, .
\ee
The quantities 
$G_z$, $G_\phi$, and $G_r$ have been neglected.
Being an odd function of $z$, $G_z$
is negligible near the midplane $z=0$, and
goes away when vertically averaging.
The fact that we are close to a purely axisymmetric
configuration (as discussed above)
implies that $G_\phi$ is also negligible for our purposes,
whereas the term containing $G_r$ in equation (\ref{pmomr})
is smaller than the other ones because radial-force balance is dominated 
by the centrifugal and gravitational terms in thin disks.

\section{CAVITY BEHAVIOR}  

In order to study the behavior of the g-mode trapping cavity 
under the inclusion of magnetic fields, 
one first obtains a dispersion relation from
the equations for the perturbations
derived in the previous subsection.
Once the characteristic equation has been obtained,
one needs to 
isolate the appropriate branch for $\omega$,
which is the one that has $\kappa$ as its leading term when the magnetic
field goes to zero.
An exploratory way of doing this is working to zeroth-order
in $k_r^2$ (i.e., setting $k_r^2 = 0$).
The curves in Fig.~1 
were obtained with this assumption, by means of a numerical
approach (using the values for $b_i$
from Table 1).
From now on we work with a non-zero $k_r^2$.

Before doing that, though, a word on dispersion relation branches.
The branches that describe
Alfv\'en waves and slow magnetosonic waves, which
are the ones responsible for the magneto-rotational 
instability (MRI),
are different ones from the one we study in this paper (Balbus \& Hawley 1998).
In particular,
the MRI branches are characterized by low frequencies
and growth rates 
of order $b_z^{1/2} \, \Omega$, which can be smaller than the 
g-mode frequencies.
More explicitly,
the typical timescale for MRI growth $\tau_{\rm MRI}$
is related to the g-mode oscillation period $\tau_{\rm g}$ by
the following formula:
\be
   \frac{\tau_{\rm MRI}}{\tau_{\rm g}} \sim \frac{\kappa}{\Omega} \, b_z^{-1/2} \, ,
\ee
which is $\sim 3$ for typical values of the involved quantities.
Even though there are MRI effects of shorter timescales ($\sim 1/\Omega$),
these only occur at very short length scales ($\ll$ disk thickness).
Furthermore, the (alpha model)
viscosity induced g-mode growth timescale $\tau_{\rm visc}$ 
(Ortega \& Wagoner 2000) is related to $\tau_{\rm MRI}$
by:
\be
    \frac{\tau_{\rm MRI}}{\tau_{\rm visc}} \sim \frac{\alpha_*}{b_z^{1/2}} \, ,
\ee
which means that the MRI might grow no faster than the viscous 
g-mode growth.
These results indicate that the g-modes may survive in the presence
of MRI driven turbulent eddies.

Another potential 
source of g-mode disruption is given by energy ``pumping''
from short to long length scales observed 
in freely decaying MHD turbulence, ``inverse-cascade'' simulations (Zrake 2014).
On closer inspection, however, it is reassuring to see that
the oscillation frequencies produced
by this mechanism at the g-mode length scales
are in reality much lower than the g-mode frequencies.

We now employ a perturbative approach in 
order to solve the problem.
Recall that
we work with the small quantities 
$\varepsilon \ll 1$ 
and $b_z^2, b_r^2 \ll b_\phi^2 \ll 1$,
and that we assume $m = 0$.

The perturbed MHD equations (\ref{wkbconta})--(\ref{wkbmhdza}),
to order $b_i b_j$,
lead to the relevant dispersion relation:
\be\label{casei}
  \omega = \omega_0 + \Lambda  b_\phi b_z + \lambda  b_\phi b_r + \Gamma  b_z^2  + \gamma  b_r^2 + \beta b_z b_r \, ,
\ee
where
\bea
\lbrack{\rm Re}(\omega_0)\rbrack^2 &=& \kappa^2  
 - \frac{\kappa^2 k_r^2 c_s^2 (1 + b_\phi^2)}{(k_z^2+k_r^2)
 c_s^2 (1 + b_\phi^2) - \kappa^2} \sim \kappa^2 \, , \\  
 {\rm Re (\Lambda)} &=&  
\frac{k_z^3 c_s^4 
[\kappa^2 + 2(1-p+q)\Omega^2]}
{2\Omega r
\kappa^2
(k_z^2 c_s^2 - \kappa^2)}
\sim \varepsilon \left( \frac{\Omega}{\kappa} \right)^2 \Omega \, , \\
 {\rm Re (\lambda)} &=&  
    \frac{ (1-p+q) c_s^2 \Omega k_r } {r \kappa^2} 
       \sim \varepsilon^{3/2} 
           \left( \frac{\Omega}{\kappa} \right)^2 \Omega \, , \\   
 {\rm Re (\Gamma)} &=& - \frac{p \Omega^2 k_z^2 c_s^2}{\kappa^3} 
\sim \left( \frac{\Omega}{\kappa} \right)^3 \Omega \, , \\
{\rm Re (\gamma)} &=& - \frac{p \Omega^2 k_r^2 c_s^2}{\kappa^3} 
   \sim \varepsilon \left( \frac{\Omega}{\kappa} \right)^3 \Omega  \, , \\
{\rm Re (\beta)} &=& \frac{2 \Omega^2 k_r k_z c_s^2}{\kappa^3} 
   \sim \varepsilon^{1/2} \left( \frac{\Omega}{\kappa} \right)^3 \Omega \, .
\eea
The leading imaginary contribution comes from the term
\be\label{imaginary}
{\rm Im (\Lambda)} =   
\frac{k_z k_r c_s^2
[2(2+p)\Omega^2-\kappa^2]}
{4\Omega (k_z^2 c_s^2 - \kappa^2)} \sim \varepsilon^{1/2} \, \Omega \, ,
\ee
the effects of which are small compared to the real-part terms.
(We note that $\Omega^2 \gg \kappa^2$ by one order of magnitude.)

We note that 
the implied inverse timescale for possible mode growth
due to equation (\ref{imaginary}) is $1/\tau \sim b_\phi b_z \varepsilon^{1/2} \Omega$,
which is much smaller than the one
corresponding to purely viscous effects (no magnetic fields) on 
an fundamental g-mode,
$1/\tau_{\rm visc} \sim \alpha_* \Omega$,
except for very small values of $\alpha_*$.

We also note that the sign of Im($\Lambda$) is not determined by our formalism
as $k_r$ and $k_z$ could have either values of the sign.

\section{DISCUSSION}

We are now in a position to offer an improved assessment of the effects
on diskoseismology
of finite magnetic fields, including the important radial magnetic fields.

Our results can be best appreciated in a plot
of the form shown in Fig.~2, which
describes the behavior of the trapping cavity in terms
of the vertical and radial magnetic fields.
The cavity is only destroyed 
(i.e., there is no value of the radius at which $d\omega(r)/dr$ 
vanishes)
outside the corresponding ellipse, for sufficiently large 
$B_z$ and $B_r$ fields.
This figure was obtained by
scanning the behavior of $\omega(r)$ for different 
values of the $B_i$ and
determining where the cavity disappears.
(The mode lives in the range of radii where $k_r^2 > 0$ for
a given eigenfrequency.)

In order to generate these results, the following ansatz 
was used: 
$k_r^2 = \varepsilon k_z^2$, which is 
consistent with a radial mode size $\sim \sqrt{hR}$
[cf.~eq.~(5.1) in Perez et al.~1997],
with $\varepsilon = 0.1$, and
$k_z^2 c_s^2 = \eta \Omega_\perp^2$ with $\eta = 1$
(as in FL), 
where 
\be
\Omega_\perp(r) = \Omega(r)(1-4a/r^{3/2}+3a^2/r^2)^{1/2}
\ee 
is the vertical epicyclic frequency.
In addition, we also used an alternative ansatz given by
$k_r^2 = \varepsilon^2 k_z^2$, corresponding to
a larger radial mode size $\sim R$, motivated by the fact 
that the g-mode radial extension might increase as
the concavity of $\omega$ decreases. (This second ansatz gives
the maximum radial g-mode extension that does not 
contradict the WKB assumption.)

Note that
the dependence on $B_\phi$ is rather weak, as long
as $B_\phi \gg B_z, B_r$.
Importantly, 
within each ellipse, 
the maximum value of $\omega(r)$ does not typically change by more than 
about $15\%$ (see Fig.~3 for a typical case), 
which means that the results are consistent
with a constant QPO frequency within the present limits of observation.
We should point out, however, that the numerical results of Fig.~1
imply a somewhat greater range of variation for the maximum value
of $\omega$,
in potential disagreement with observations.
(Recall, though, that these results assume $k_r^2 = 0$.)

Even though 
the perturbative results cannot be directly compared to FL
(who study only the 
$\mathbf{B} = B_\phi(r) \uvp$ and
$\mathbf{B} = B_z \uvz$ special cases, separately),
their results are consistent with ours
in general terms.

Our main conclusions 
in the present exploratory approach 
are the following.
First, from the above discussion
there seems to be no compelling reason to discard 
axisymmetric g-mode 
$\kappa$ trapping. 
While it is still true 
that the inclusion of magnetic fields modifies the cavity,
the situation is not as devastating as
implied by FL.
Note in particular that
the inclusion of a non-zero $B_r$ potentially allows for 
slightly larger values of cavity-preserving $|B_z|$.
Furthermore,
the numerical results of Fig.~1 hint that the perturbative
results may be underestimates of these $|B_z|$ values.

More importantly, most simulations
appear to produce values of $B_r$ and $B_z$ 
which lie within
or near
each ellipse of Fig.~2. See Table 1 and corresponding bullets in Fig.~2.
(Note, however, that there is an outlier point, not plotted.)

In the second place, it must be kept in mind 
that possible diskoseismic explanations of QPOs require
only that the magnetic field be inside the ellipses in Fig.~2
during some, possibly small, fraction of the time, 
as the corresponding QPO duty cycles are 
observed to be much smaller than 100\% (Remillard \& McClintock 2006; 
Belloni, Sanna, \& M\'endez 2012). 

\acknowledgments
This work was supported by 
grant 829-A3-078 of 
Universidad de Costa Rica's Vicerrector\'{\i}a de Investigaci\'on
and by grant FI-0204-2012 of MICITT and CONICIT.
Travel funds provided by Stanford and Universidad de Costa Rica.

\newpage

\begin{deluxetable}{llcccc}
\tablewidth{0pt}
\tablecaption{Saturation Values of the Magnetic Field.}
\tablehead{
\colhead{Type of 3D Simulation} & \colhead{Initial $\mathbf{B}$ Field} & 
\colhead{$|b_z|$} & 
\colhead{$|b_r|$} &
\colhead{$|b_\phi|$} & 
\colhead{Reference} }
\startdata
shearing box, purely radial gravity & vertical & 
\phantom{$< \,$}0.09 & \phantom{$< \,$}0.15 & \phantom{$< \,$}0.26 & 1 \\ 
shearing box, purely radial gravity  & toroidal & 
\phantom{$< \,$}0.04 & \phantom{$< \,$}0.07 & \phantom{$< \,$}0.21 & 1 \\ 
shearing box, purely radial gravity  & $\langle\mathbf{B}\rangle=0$ & 
$< \,$0.03  & $< \,$0.05 & $< \,$0.16 & 2 \\ 
shearing box, including vertical gravity  & twisted toroidal & 
\phantom{$< \,$}0.09  & \phantom{$< \,$}0.11 & \phantom{$< \,$}0.28 & 3 \\  
shearing box, including vertical gravity  & vertical & 
\phantom{$< \,$}0.06  & \phantom{$< \,$}0.08 & \phantom{$< \,$}0.23 & 4 \\ 
global, magnetically choked, $H/R \approx 0.5$ & vertical & 
\phantom{$< \,$}0.08 & \phantom{$< \,$}0.56 & \phantom{$< \,$}0.56 &  5 \\ 
global, $H/R = 0.1$ & toroidal  & 
$< \,$0.05 & $< \,$0.07 & $< \,$0.17 & 6 \\ 
global, different temperature profiles  & vertical, weak  & 
$< \,$0.04 & $< \,$0.06 & $< \,$0.17 & 7 \\  
\enddata
\tablecomments{Saturation values of the magnitude of the magnetic field components 
are quoted, 
in the form of $b_i \equiv v_{Ai}/c_s$,
according to various 3D simulations. $H/R$ refers to the 
the disk's typical thickness to radius ratio.}
\tablerefs{
(1) Hawley et al.~1995;
(2) Hawley et al.~1996;
(3) Shi et al.~2010;
(4) Simon et al.~2011; 
(5) McKinney et al.~2012;
(6) Parkin 2013;
(7) Suzuki \& Inutsuka 2014.
}
\end{deluxetable}

\clearpage

\begin{figure}
\epsscale{.58}
\plotone{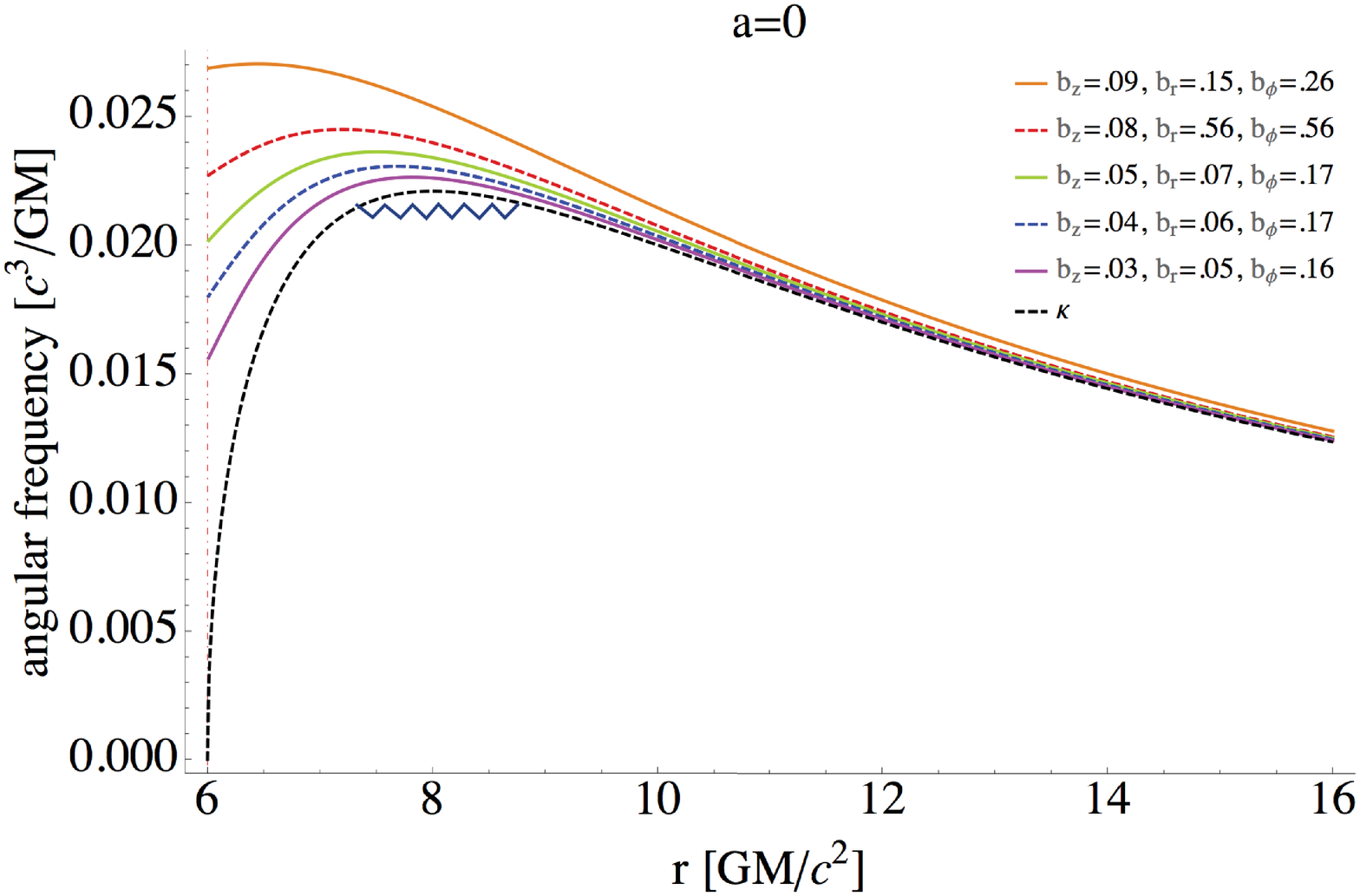}
\end{figure}

\begin{figure}
\epsscale{.58}
\plotone{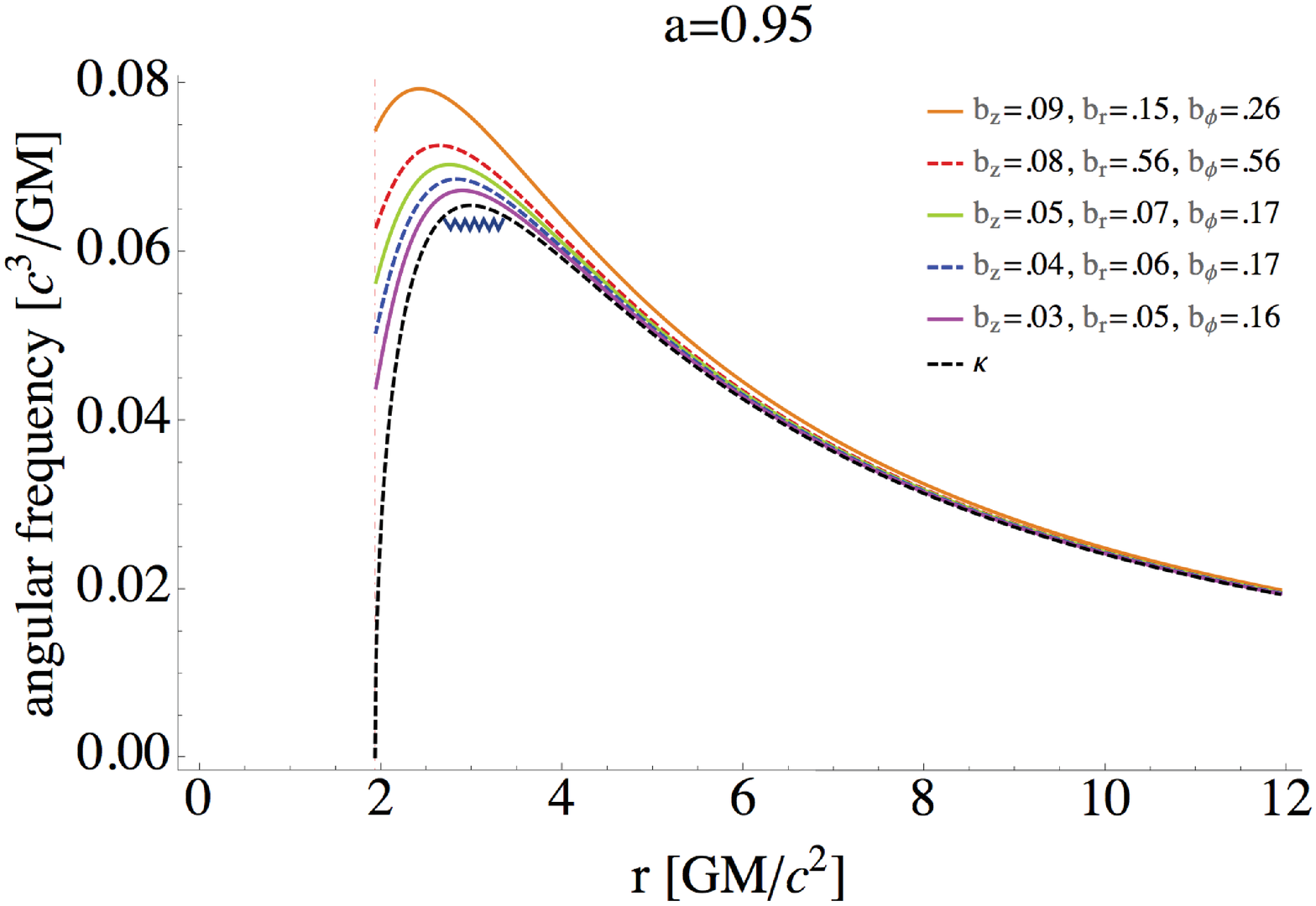}
\caption{g-mode trapping cavity numerical estimates
for different values of the magnetic field,
assuming $m=0$ and setting $k_r = 0$ as
a first approximation (finite $k_r$ results within a 
perturbative approach are presented in Fig.~2 and discussed in Section 5).
The normalized magnetic fields $b_i$ are defined in equation (\ref{defbi}),
and their values are taken from Table 1.
(The listed $b_i$ values at the upper right corners correspond 
to the curves, in descending order.)
For a given frequency, oscillating modes can exist below the
corresponding curve.
For high enough values of the magnetic field, the cavity is destroyed
(not shown), i.e., the curve fails to have a maximum.
Also shown is $\kappa(r)$, the leading term of the cavity in the absence of magnetic
fields.
Jagged curves are g-modes (shown here for the case of vanishing magnetic fields),
dash-vertical lines mark the inner disk boundary at the ISCO.
The upper and lower panels are for the respective cases of 
$a=0$ and $a=0.95$, 
where $a \equiv cJ/GM^2$ is the black hole angular momentum parameter.
(This figure appears in color in the online version of the paper.)
}
\end{figure}

\newpage

\begin{figure}
\epsscale{.58}
\plotone{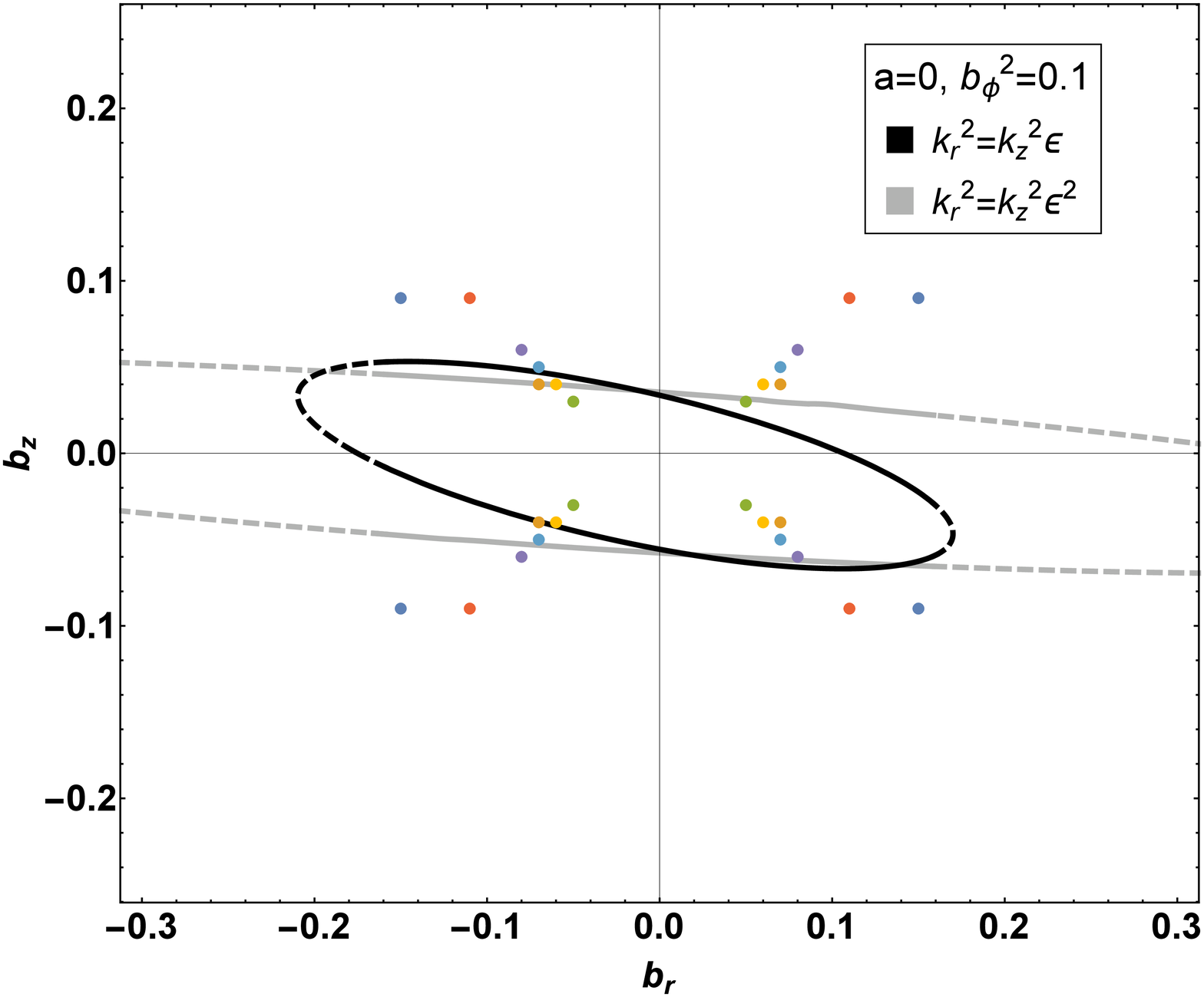}
\end{figure}

\begin{figure}
\epsscale{.58}
\plotone{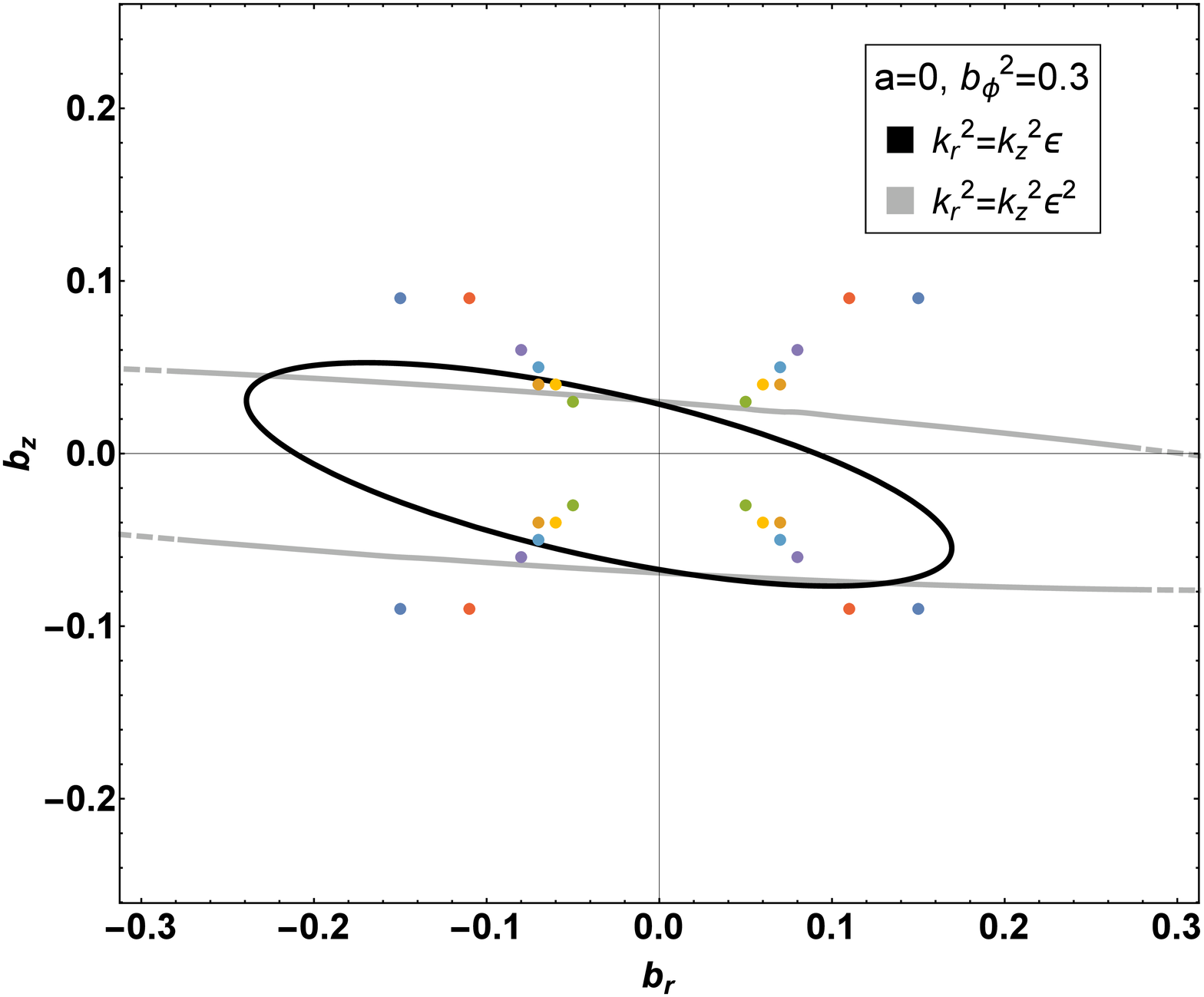}
\end{figure}

\begin{figure}
\epsscale{.58}
\plotone{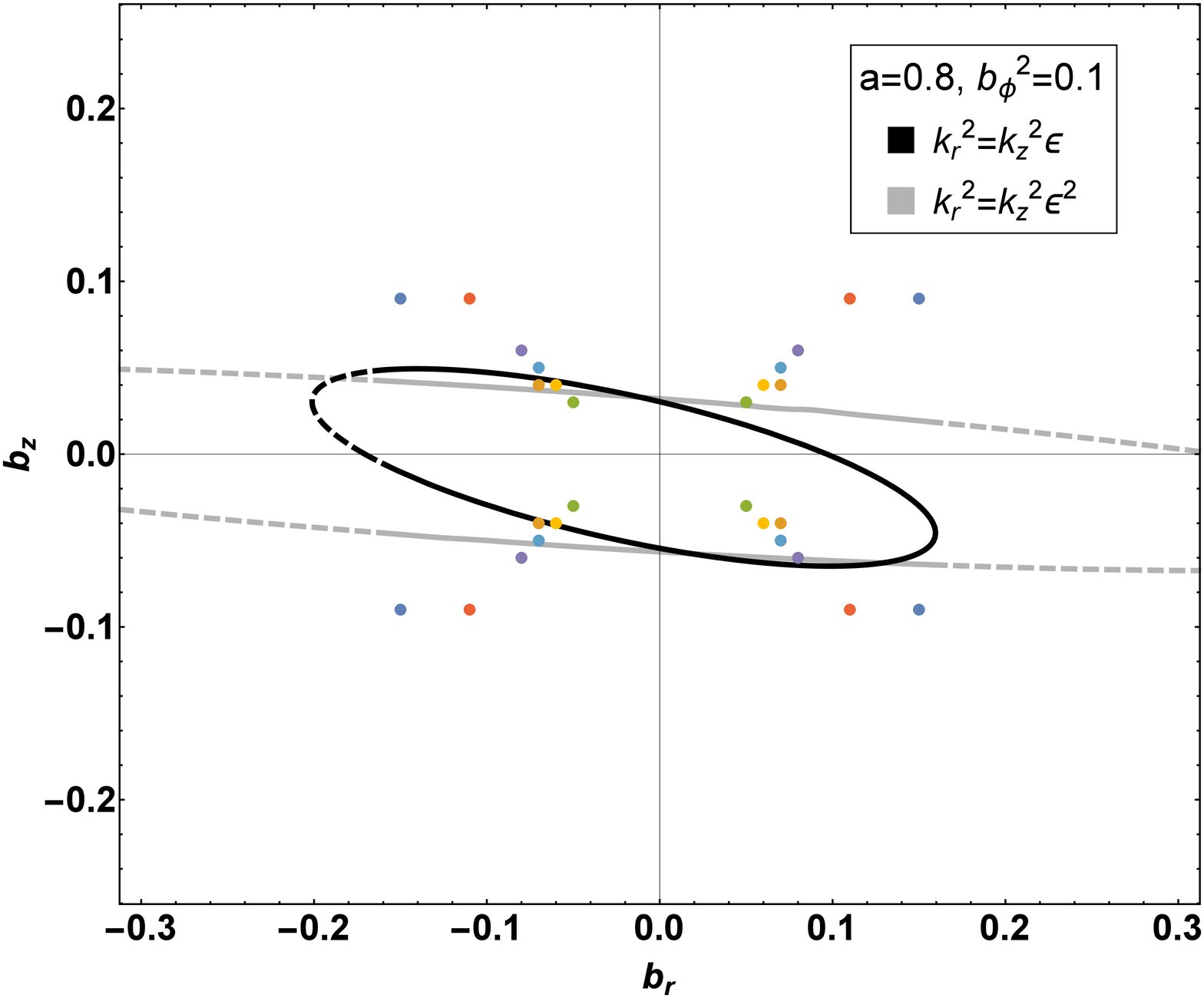}
\end{figure}

\begin{figure}
\epsscale{.58}
\plotone{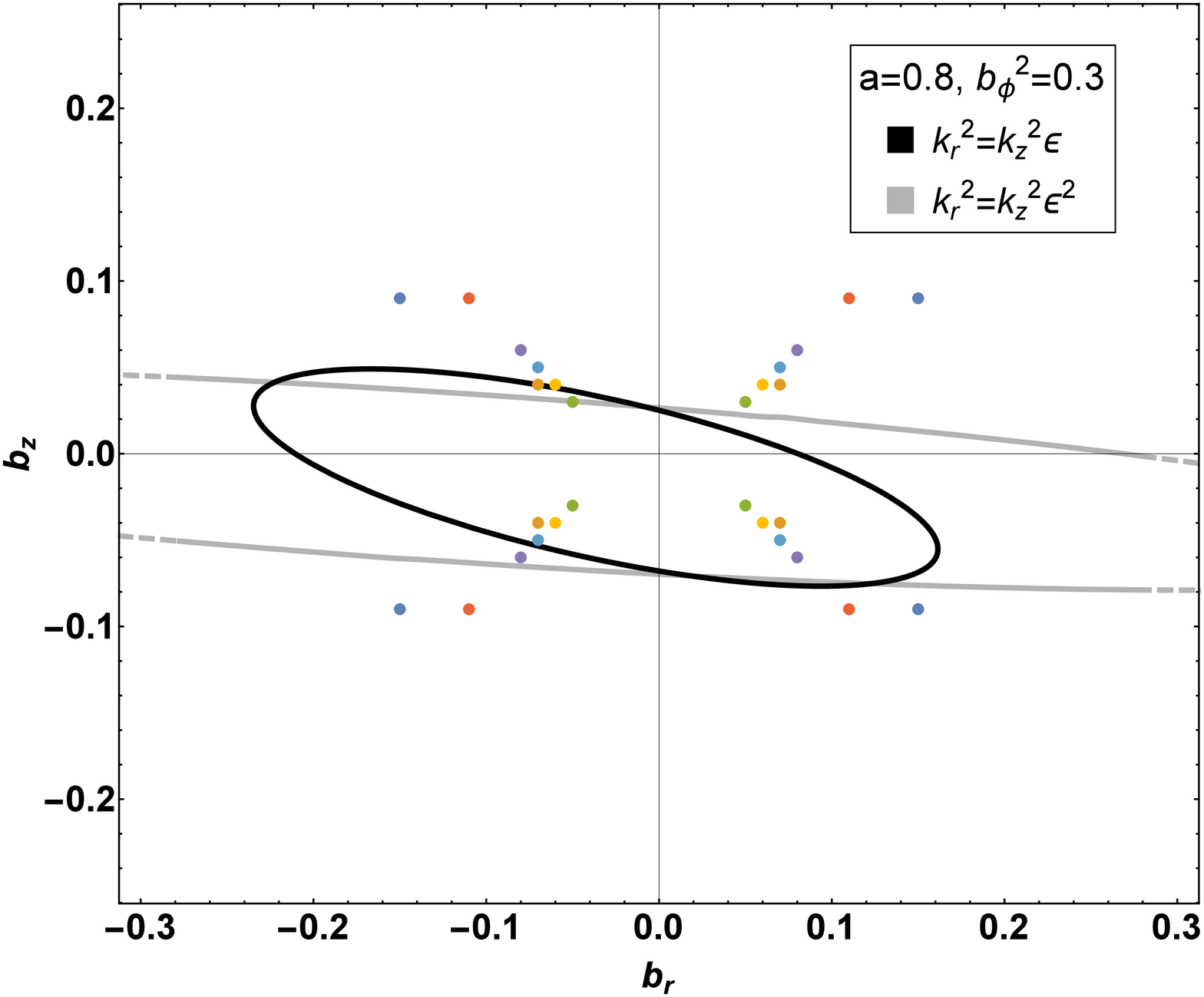}
\caption{Behavior of the g-mode trapping cavity as a function of the three 
(normalized) components
of the magnetic field $b_i$ for different values of $a$.
The cavity is preserved within each ellipse
(dark and light for 
$k_r^2 = k_z^2 \varepsilon$ and $k_r^2 = k_z^2 \varepsilon^2$,
respectively, corresponding to different radial mode sizes), 
and destroyed outside it.
Extrapolations to perturbative analysis are indicated by dashes;
they occur whenever the maximum of $b_r^2$ and $b_z^2$ is 
larger than $b_\phi^2/4$.
Also shown are bullets corresponding to Table 1 simulation 
saturation values (or their upper bounds), but note that we plot $\pm$ the values,
as they carry no sign. We do not plot the outlier point.
(This figure appears in color in the online version of the paper.)
}
\end{figure}

\newpage

\begin{figure}
\epsscale{.6}
\plotone{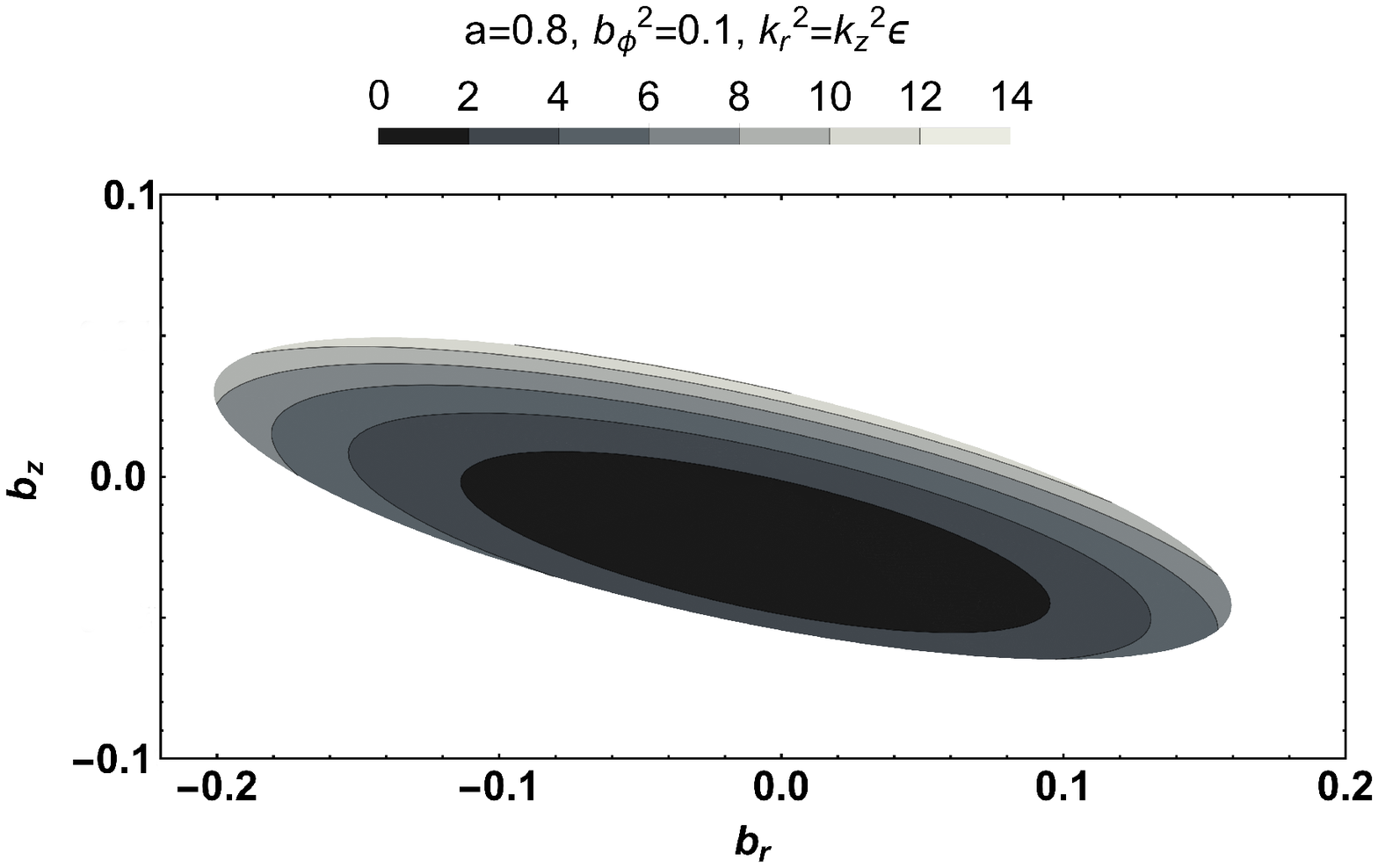}
\caption{Behavior of the variation of the maximum value
of $\omega$ within the (perturbative analysis) trapping-cavity ellipse for the case
$a=0.8$, $b_\phi^2=0.1$, $k_r^2 = k_z^2 \varepsilon$ in Fig.~2, 
as a function of $b_z$ and $b_r$.
Different shades of gray represent percentual differences,
from 0\% to 14\%, with respect to the smallest maximum value.
(See, however, the comment in the main text about the implications of 
the numerical results of Fig.~1.)}
\end{figure}


\begin{references}

\reference{} Balbus, S. A., \& Hawley, J. F. 1998,
             Rev. Mod. Phys., 70, 1

\reference{} Belloni, T. M., Sanna, A., \& M\'endez, M. 2012,
             MNRAS, 426, 1701

\reference{} Fu, W., \& Lai, D. 2009, ApJ, 690, 1386 (FL)

\reference{} Hawley, J. F., Gammie, C. F., \& Balbus, S. A. 1995, ApJ, 440, 742

\reference{} Hawley, J. F., Gammie, C. F., \& Balbus, S. A. 1996, ApJ, 464, 690

\reference{} McKinney, J. C., Tchekhovskoy, A., \& Blandford, R. D. 2012, 
             MNRAS, 423, 3083

\reference{} Novikov, I. D., \& Thorne, K. S. 1973, in 
             Black Holes, ed. C. DeWitt \& B. S. DeWitt 
             (New York : Gordon \& Breach), 343

\reference{} O'Neill S. M., Reynolds C. S., \& Miller C. M. 2009, ApJ, 693, 1100 

\reference{} Ortega-Rodr\1guez, M., \& Wagoner, R. V. 2000, ApJ, 537, 922

\reference{} Parkin, E. R. 2014, MNRAS, 441, 2078

\reference{} Perez, C. A., Silbergleit, A. S., Wagoner, R. V., \& Lehr, D. E.
             1997, ApJ, 476, 589

\reference{} Remillard, R. A., \& McClintock, J. E. 2006,  
             Annu. Rev. Astron. Astrophys., 44, 49
 
\reference{} Reynolds, S., \& Miller, M.C. 2009, ApJ, 692, 869

\reference{} Shi, J., Krolik, J. H., \& Hirose, S. 2010, ApJ, 708, 1716

\reference{} Simon, J. B., Hawley, J. F., \& Beckwith, K. 2011,
             ApJ, 730, 94

\reference{} Suzuki, T. K., \& Inutsuka, S. I. 2014, ApJ, 784, 121

\reference{} Wagoner, R. V. 2008, New Astronomy Reviews, 51, 828

\reference{} Zrake, J. 2014, ApJ, 794, L26

\end{references}
\end{document}